\documentstyle[preprint,aps]{revtex}
\begin{document}
\draft
\title{\bf Exact solutions in two--dimensional string
cosmology \\ with back reaction}
\author{Sukanta Bose \thanks{Electronic Address :
sbose@iucaa.ernet.in} and Sayan Kar \thanks{Electronic Address :
sayan@iucaa.ernet.in}} 
\address{Inter--University Centre for Astronomy and Astrophysics,\\
Post Bag 4, Ganeshkhind, Pune, 411 007, INDIA}
\maketitle
\widetext
\parshape=1 0.75in 5.5in
\begin{abstract}

We present analytic cosmological solutions in a model of two-dimensional 
dilaton gravity with back reaction. One of these solutions exhibits a 
graceful exit from the inflationary to the FRW phase and is nonsingular
everywhere. A duality related second solution is found to exist only in the 
``pre-big-bang'' epoch and is singular at $\tau = 0$. In either case
back reaction is shown to play a crucial role in determining the specific
nature of these geometries.
 
\end{abstract}
\vskip 0.125 in
\parshape=1 0.75in 5.5in
\newpage

Two-dimensional (2D) models of gravity have acquired a special status over
the last few years. In spite of being largely pedagogical, they form
a useful set where the usual problems of four-dimensional (4D) theories of
gravity such as black--hole evaporation, cosmological singularities, etc.
can be addressed and answered with reasonable confidence. 
Thus, one expects that the physical insights obtained in such a 2D
setting would carry over to some extent to the actual 4D 
scenario, although the technicalities will be more involved there.

Among such models, the one that has been investigated quite thoroughly, is 
due to Callan, Giddings, Harvey, and Strominger (CGHS) \cite{CGHS}. The CGHS 
model is largely inspired by target space, low--energy effective string theory
compactified down to two dimensions.
The field content of this model constitutes the dilaton ($\phi$), the graviton
($g_{\mu\nu}$), and a multicomponent conformally coupled massless 
scalar matter field denoted by $f_{i}$ ($i=1,2,.....N$). The CGHS action is
\begin{equation}
\label{CGHS}
S_0 = {1\over 2\pi} \int d^{2}x \sqrt{-g}\left \{ e^{-2\phi}\left[ R +
4\left (\nabla \phi
\right ) ^{2} - 4\Lambda \right ] - \frac{1}{2}\sum_{i=1}^{N}
\left ( \nabla f_{i} \right )^{2} \right \} 
\ \ ,
\end{equation}
where $\Lambda$ is a cosmological constant term, which will henceforth be
set equal to zero. 

The cosmological solutions to the above 2D action were first studied by
Mazzitelli and Russo \cite{MazRus}. However, it is the recent work of
Rey \cite{Rey} that has revived interest in this model. Rey showed that the 
above action has classical solutions with features similar to
the cosmological solutions in 4D superstring theory (for reviews, see Refs. 
\cite{Ven-rev}). These solutions are characterised by two 
branches 
separated in time by a curvature singularity. The first branch describes 
superinflationary behavior, with accelerated expansion and monotonically
increasing scalar curvature, whereas the second branch describes an FRW 
universe. Although the existence of a superinflationary phase in the early
history of the universe is promising, the full solution is plagued by the 
lack of a smooth singularity-free transition from this ``pre-big-bang'' phase
to the FRW phase. This is known as the ``graceful exit'' problem of 
superstring cosmology.

It was also shown by Rey that incorporating back reaction effects cures the
graceful exit problem. (For one approach towards a solution to the back 
reaction problem in two dimensions, see Ref. \cite{NS}.) However, his cure 
depended on the restrictive requirement that the total number, $N$, of 
massless scalar matter fields in the theory be less than twenty four. 
Subsequently, there have been some attempts 
at removing this barrier by modifying the 2D classical action and studying 
the back reaction effects on corresponding solutions \cite{GaspVen,KimYoon}.
However, in these papers, in cases where the restriction on $N$ is removed,
one finds that, unlike the CGHS action, the corresponding 2D classical actions
cannot be motivated by reduction from a realistic 4D action.

It was shown in \cite{Bose} that one can remove the barrier on $N$ in Rey's 
model without 
modifying the tree level action (\ref{CGHS}). As in Rey, the main idea 
in \cite{Bose} was to study back reaction effects by incorporating one-loop 
correction terms in the CGHS action. However, since these one-loop terms
are well defined only up to the addition of local covariant counterterms 
\cite{OSRusTseyt}, different combinations of such terms lead to different 
models. One chooses to analyse only those models that display the physical 
phenomena of interest. The counterterms used in the one-loop action of 
\cite{Bose} are different from those used by Rey (which were the ones
corresponding to the RST model \cite{RST}). The resulting quantum-corrected
cosmological solutions of \cite{Bose} are similar to Rey's, with the added 
advantage that the graceful exit problem is solved for any positive 
value of $N$.

It is important to note that only the asymptotic behavior of the 
quantum-corrected solutions were studied in Refs. \cite{Rey} and \cite{Bose}.
In this paper, we find the exact solutions of these models and discuss their
physical features. In particular, we address the following two issues. We 
show that in our quantum-corrected solutions, where the graceful exit problem
is solved, the Weak Energy Condition is violated. We also show that the scale 
factor duality is not respected by the exact solutions. 
However, there exists a duality transformation that relates one solution to
another.

The one-loop corrected 2D action studied in Ref. \cite{Bose} is
\begin{equation}
S_1 = S_0 + {N\hbar \over 24\pi} \int d^2 x  \sqrt{- g} \>
( - {1\over 4} R \Box^{-1} R + 2(\nabla \phi)^2 - 3 \phi R )
\ \ ,
\label{1loopac}
\end{equation}
where $\Box_x G(x, x') =  \delta^2 (x-x') / \sqrt{-g(x)}$. The first term in 
the parenthesis is the Polyakov-Liouville term that reproduces the trace 
anomaly for massless scalar fields \cite{CGHS,BD}. The remaining terms in the
parenthesis are local covariant counterterms that differ from the ones 
occurring in the RST model. The higher order corrections beyond one loop are 
dropped by using the large $N$ approximation where $N \to \infty$ as $\hbar 
\to 0$ such that $\kappa \equiv N\hbar / 12$ remains finite. 
The metric components in the conformal gauge, $g_{\mu\nu} = e^{2\rho} 
\eta_{\mu\nu}$, when expressed in the double null coordinates, $x^\pm \equiv 
t \pm x$, are $g_{+-} = -e^{2\rho} /2$ and $g_{\pm\pm} = 0$.

We now use the following redefined fields
\begin{equation}
\label{qPS}
\Sigma \equiv (\phi - \rho) , \hskip1cm 
\Phi \equiv e^{-2 \phi} - \kappa\phi + {\kappa \over 2} \rho
       = e^{-2 \phi} - {\kappa \over 2} \rho -\kappa\Sigma  
\ \ .
\end{equation}
In terms of these fields the one-loop corrected action in the conformal 
gauge simplifies to
\begin{equation}
\label{qacconf}
S_1 = {1\over \pi} \int d^2 x \left[ \partial_+ \Phi \partial_- \Sigma
+ \partial_- \Phi \partial_+ \Sigma  + {1\over 2} \sum_{i=1}^N
\partial_+ f_i \partial_- f_i \right] 
\ \ ,
\end{equation}
where we have set $\Lambda = 0$.
For homogeneous cosmologies with constant $f_i$'s, the equations of motion 
following from the variation of the above action are 
\begin{equation}
\label{redvareom}
{d^2 \Phi \over dt^2} = 0 = {d^2 \Sigma \over dt^2}
\,.
\end{equation}
The accompanying constraints (obtained from varying $S_1$ in
(\ref{1loopac}) with respect to $g^{\mu\nu}$ and setting $\mu=\nu=\pm$) are
\begin{equation}
\partial_{\pm}^2 {\Phi} + 2\partial_{\pm} {\Phi}\partial_{\pm} \Sigma 
= {3\over 2}\kappa \left[ \partial_{\pm}^2
\phi - 2\partial_{\pm} {\rho} \partial_{\pm} {\phi} \right] 
+\kappa t_{\pm} (x^{\pm}) 
\ \ ,
\label{qcons}
\end{equation}
where $t_{\pm} (x^{\pm})$ are nonlocal functions that arise 
from the homogeneous part of the Green function (see \cite{CGHS,BPP}). 
The choice of these nonlocal 
functions determine the quantum state of the matter fields in the spacetime.
The total matter stress tensor can be expanded in orders of $\hbar$ as
$T^f_{\mu \nu} = (T^f_{\mu \nu})_{c\ell} + \langle T_{\mu \nu} \rangle$,
where $(T^f_{\pm \pm})_{c\ell} \equiv {1\over 2}\sum_{i=1}^N (\partial_{\pm} 
f_i )^2 $ is the classical part and $\langle T_{\pm \pm} \rangle = 
\kappa [ \partial^2_{\pm} \rho - (\partial_{\pm} \rho)^2 - t_{\pm}(x^{\pm})]$ 
is the one-loop contribution \cite{DFU,BD}. 
We will choose the quantum state of the matter fields to be defined by
\begin{equation}
\label{qt}
t_{\pm} (x^{\pm}) = - {3\over 2} \left[ \partial_{\pm}^2
\phi - 2\partial_{\pm} {\rho} \partial_{\pm} {\phi} \right] \ \ ,
\end{equation}
which simplifies the constraint equations for the homogeneous cosmologies to 
\begin{equation}
\label{homcons}
{d^2 {\Phi} \over dt^2} + 2 {d {\Phi}\over dt}{d {\Sigma}\over dt} = 0
\,.
\end{equation}
The equations of motion (\ref{redvareom}) and the above constraint have been
solved to obtain the past and future asymptotic behavior of cosmological 
solutions \cite{Bose}. 

To obtain the exact expression for the scale factor $a$ in these solutions, 
we define $a(t) \equiv \ln \rho(t)$ where, as defined above, $t$ is the 
conformal time. We then rewrite the field equations and the constraint in 
terms of $\phi(\tau)$ and $a(\tau)$, where $\tau$ is defined to be the 
comoving time obeying $d\tau /a(\tau) \equiv dt$. They turn out to be:
\begin{equation}
\label{eom1}
\ddot{\phi} - \frac{\ddot a}{a} + \frac{\dot a}{a} \dot \phi = 0 \>,
\end{equation}
\begin{equation}
\label{eom2}
2e^{-2\phi} \left ( 2{\dot\phi}^{2} - \ddot \phi
- \frac{\dot a}{a} \dot \phi \right ) + \frac{1}{2}\kappa
\left ( -2\frac{\dot a}{a}\dot \phi + \frac{\ddot a}{a} - 2\ddot \phi \right ) 
= 0 \>,
\end{equation}
\begin{equation}
\label{eom3}
\left ( -2 e^{-2\phi}\dot \phi - \kappa \dot \phi + 
\frac{\kappa}{2}\frac{\dot a}{a} \right ) \left ( \dot \phi - \frac{\dot a}{a}
\right ) = 0 \>,
\end{equation}
where the overdot denotes $d/d\tau$.
The first two equations in this set are the equations of motion and the third
is the constraint. 

Note that in the classical case ($\kappa = 0$), the above equations are 
invariant under the scale factor duality (SFD) transformations:
\begin{equation}
\label{SFD}
\phi \to \phi - \ln a \qquad , \qquad a \to 1/a
\,.
\end{equation}
In terms of the redefined fields (\ref{qPS}) (with $\kappa =0$), this duality 
is essentially related to the invariance of the set of equations 
(\ref{redvareom}) and (\ref{homcons}) under the transformation $\Phi \to 
\Sigma$ and $\Sigma \to \Phi$. Since such a transformation leaves the quantum
equations ($\kappa \neq 0$) invariant as well, there should exist a duality 
transformation analogous to (\ref{SFD}) for the one-loop corrected equations.
We find that Eqs. (\ref{redvareom}) and (\ref{homcons}) are 
invariant under the dual transformation:
\begin{equation}
\label{qDT}
\phi \to \phi \qquad , \qquad \ln a(t) \to -\frac{2}{\kappa}e^{-2\phi} 
          - \ln a(t) + 3\phi
\ \ ,
\end{equation}
which is equivalent to the interchange of $\Phi$ and $2\Sigma /\kappa$.
Such a symmetry can be exploited to construct both classical as well as
quantum solutions. A similar symmetry ($\phi \to \phi$, $\ln a(t) \to 
-\frac{2}{\kappa}e^{-2\phi} - \ln a(t) + \phi$) exists in Rey's model 
as well.

We now obtain the solutions in our one-loop corrected model.
The constraint (\ref{eom3}) contains two factors. We first choose to solve it
by setting
\begin{equation}
\dot \phi = \frac{\dot a}{a} \qquad {\rm or} \qquad \phi = \ln a
\,.
\end{equation}
Note that the first equation (\ref{eom1}) is also solved by this condition 
on $\phi$. Substituting the expressions for $\phi$ and its derivatives 
into (\ref{eom2}), we obtain the following equation for the scale factor 
$a(\tau)$:
\begin{equation}
\label{a-eom}
\frac{2}{a^{2}} \left [ 2 \left (\frac{\dot a}{a}\right )^{2}
-\frac{\ddot a}{a} \right ] - \frac{\kappa}{2} \frac{\ddot a}{a} = 0
\,.
\end{equation}
Note that for $a$ large the first term is small and we get a solution that 
is linear in $\tau$. For $a$ small, on the other hand, it is the
first term that dominates and we obtain the superinflationary
epoch in the asymptotic past, where the scale factor exhibits an inverse
power-law behavior in $\tau$. Thus, scale factor duality symmetry
is preserved only in the asymptotic past and future. 

We solve Eq. (\ref{a-eom}) by rewriting it as:
\begin{equation}
\frac{d^{2}}{d\tau^{2}} \left ( -\frac{1}{a} + \frac{\kappa}{4} a
\right )  = 0
\ \ ,
\end{equation}
which yields
\begin{equation}
\label{a1st}
a(\tau) = \frac{2}{\kappa} \left ( \alpha \tau + \beta \pm
\sqrt{\left(\alpha \tau + \beta \right )^{2} + \kappa}\right )
\ \ ,
\end{equation}
where $\alpha$ and $\beta$ are integration constants, and $\kappa$ is 
positive.
The solution with the negative sign for the square root has to be discarded 
because it gives a negative $a$. One can check that in the asymptotic past 
and future, the scale factor is proportional to $-\frac{1}{\tau}$ and $\tau$, 
respectively. Hence, this solution exhibits scale factor duality 
asymptotically. The function (with the $+$ sign for the square root) is 
plotted in Fig. 1. Solution (\ref{a1st}) is the quantum analogue of the 
first branch solution \cite{Rey} in the classical case.

The important feature of this solution is that $\kappa$ plays a
crucial role in it. Without a finite value of $\kappa$ one would
end up with a scale factor that either vanishes or diverges for a finite
value of $\tau$. This would preclude
a solution to either the graceful exit problem or the singularity problem. 

The Ricci scalar for the above geometry turns out to be:
\begin{equation}
R = \frac{2\alpha^{2} \kappa}{\left \{ \left ( \alpha \tau + \beta \right )
^{2} + \kappa \right \}^{\frac{3}{2}} 
\left \{ \alpha
 \tau + \beta + \sqrt{\left (\alpha
\tau + \beta \right )^{2} + \kappa}
 \right \}}
\ \ ,
\end{equation}
which is everywhere finite, clearly due to a finite value of $\kappa$.
 
It is instructive to construct the exact solution of the field equations and
constraints for the model due to Rey \cite{Rey}. Following the steps similar 
to Eqs. (\ref{eom1})-(\ref{eom3}), we find the solution 
to be:
\begin{equation}
a(\tau) = \frac{2}{\kappa}\left \{-\left (\alpha \tau + \beta \right ) \pm
\sqrt{\left (\alpha \tau + \beta \right )^{2} - \kappa} \right \}
\,.
\end{equation}
However, here it is necessary to consider the negative sign for the square
root and also assume $\kappa$ to be negative. Substituting $\kappa =
-\vert \kappa \vert $ we get,

\begin{equation}
\label{a1rey}
a(\tau) = \frac{2}{\vert\kappa \vert}
\left \{\left (\alpha \tau + \beta \right ) +
\sqrt{\left (\alpha \tau + \beta \right )^{2} + \vert\kappa \vert} \right \}
\ \ ,
\end{equation}
which is exactly the same as the solution for the previous model except that 
we need to choose $\kappa$ to be negative. In fact, it is also easy to see 
that here the geometry is nonsingular only if $\kappa$ is negative. 
In the above solutions for both these models, note that the dilaton
field coupling, $e^{\phi}$, is actually equal to the scale factor $a$
and, therefore, remains finite except for $\tau \to \infty$ (see Fig. 1).
  
Equations (\ref{a1st}) and (\ref{a1rey}) for the scale factor in the 
first branch show that SFD is not respected in the
exact solutions of the quantum-corrected models of Refs. \cite{Rey} and 
\cite{Bose}, although it is exhibited asymptotically as $\tau \to \pm \infty$.
Even without explicitly solving the equations for $a(\tau)$ one could
deduce qualitatively the nature of the solutions from them.
Probable solutions with a graceful exit can be categorized  
into three classes: (i) $a(\tau)$ is a monotonic function,
(ii) $a(\tau)$ has atleast a pair of extrema (a maximum and a minimum)
in the neighborhood of $\tau =0$, (iii) $a(\tau)$ has inflection points
but no extrema. 

For case (ii), with $\dot a =0$ at the extrema, we must have from (16),   

\begin{equation}
\label{a0-eom}
-\frac{\ddot{a} (\tau_0 )}{a_0} \left[ \frac{2}{a_0^{2}} + \frac{\kappa}{2} 
\right] = 0
\,.
\end{equation}
at the extrema. Note that with $\kappa$ replaced by
$\vert \kappa \vert$, we get the corresponding condition for Rey's model.   
In the classical case ($\kappa =0$), the above equation can be satisfied 
if either $a_0 \to \infty$ or $\ddot{a} (\tau_0 ) =0$. The latter condition
would imply a vanishing scalar curvature for a finite $a_0$. 
The classical equations of motion in fact yield two solutions, related by 
SFD, such that the first branch (superinflationary phase) satisfies 
$a_0 \to \infty$ (as $\tau_0 \to 0$), and the second branch (FRW phase)
has $\ddot{a} (\tau_0 ) =0$. Conditions (i) and (iii) are not realised in
the classical case. Thus SFD precludes a solution to the graceful exit
problem in the classical case.

In the quantum case $\kappa \neq 0$, and the equations of motion show that 
SFD is lost (although the duality transformation (\ref{qDT})
exists that allows one to generate the second branch solution from the 
first). For case (ii) we need $a_{0}^{2}$ to be negative, which is
physically meaningless. Cases (i) and (iii) can not
be ruled out by qualitative reasoning. The exact solutions however
exhibit that the scale factor in the first branch is a function satisfying 
(i) with asymptotic SFD. Hence, the first branch itself is an interesting 
cosmological solution without the graceful exit problem.

The constraint equation (\ref{eom3}) (and a corresponding equation for Rey's
model) contains a couple of factors, one of which is common to both the 
models, namely, $(\dot{\phi} -\dot{a} /a)$. In the previous analysis, we have 
set this factor to zero and therefore the corresponding solutions in 
the two models are the same, modulo the sign of $\kappa$. We now investigate 
the consequences of setting the other factor to zero in each model.

For the model defined in (\ref{1loopac}), the constraint equation 
(\ref{eom3}) is solved by the relation,

\begin{equation}
-2e^{-2\phi}\dot \phi -\kappa\dot \phi +\frac{\kappa}{2}
\frac{\dot a}{a} = 0
\,.
\end{equation}
Defining $\xi \equiv e^{-2\phi}$, the above condition yields
\begin{equation}
\dot \xi + \frac{\kappa}{2}\frac{\dot \xi}{\xi} + \frac{\kappa}{2}
\frac{\dot a}{a} = 0
\ \ ,
\end{equation}
which gives:

\begin{equation}
\label{a2}
a = \frac{1}{\xi}e^{-\frac{2}{\kappa}\xi}
\,.
\end{equation}
The two equations of motion (\ref{eom1}) and (\ref{eom2}) reduce to the 
following differential equations in $\xi$ and its derivatives,

\begin{equation}
\left ( 1+\frac{\kappa}{4\xi}\right ) \ddot \xi - \left (\frac{3}{2\xi}
+\frac{2}{\kappa} + \frac{\kappa}{2\xi^{2}} \right ) {\dot\xi}^{2} = 0
\ \ ,
\end{equation}

\begin{equation}
\label{br2eom}
\left ( \frac{2}{\kappa}+\frac{1}{\xi} \right )
\left [ \left ( 1+\frac{\kappa}{4\xi}\right ) \ddot \xi - \left (\frac{3}
{2\xi} +\frac{2}{\kappa} + \frac{\kappa}{2\xi^{2}} \right ) {\dot\xi}^{2} 
\right ] = 0
\,.
\end{equation}
Notice that the prefactor in the second equation of motion, if set to
zero, results in a constant $\xi$ (and therefore a constant $\phi$).
Discarding this trivial solution, we finally have only one
differential equation to solve, namely, Eq. (\ref{br2eom}).

The solution to this equation gives us $\tau$ as a function of $\xi$ :

\begin{equation}
\label{tau2}
\tau = \kappa \left [ -2Ei(-2\frac{\xi}{\kappa}) -
\kappa\frac{e^{-2\frac{\xi}{\kappa}}}{\xi} \right ]
\end{equation}
where $Ei(-\xi)$ is the exponential integral function \footnote{We follow
the definition, $Ei(-x) = -\int_x^{\infty} {e^{-x} \over x} dx$.}.

Figures 2(a) and 2(b) illustrate the two functions $\tau (\xi)$ and 
$a(\tau )$, respectively. Note that only negative values of $\tau$ are 
allowed because otherwise $\xi$ is negative, which implies a negative
coupling to gravity.
The solution for $a(\tau)$ has a singularity at $\tau =0$ and there is
no physically meaningful solution in the $\tau >0$ region. It can be 
verified that the second branch solution for $a$, given by (\ref{a2}) and 
(\ref{tau2}), is related to the first branch solution (\ref{a1st}) through 
the duality transformation (\ref{qDT}).

In Rey's model, the solution corresponding to the second
branch is not expressible in terms of known functions, in
the region of positive dilaton coupling.
 
A comment about the {\em energy conditions} in these geometries.
In $1+1$ dimensions, the Raychaudhuri equation for timelike
geodesic congruences is given as \cite{CM,SK}:

\begin{equation}
\frac{d\theta}{d\lambda} + \theta^{2} = \frac{1}{2} R
\,.
\end{equation}
The condition for focusing of timelike geodesics therefore reduces to
$R\le 0$. 
For the cosmological metrics under consideration $R = 2\ddot{a} / a$.
It is therefore clear that geodesics will not focus in this geometry
within a finite value of $\lambda$ because the convergence 
condition is violated. In fact the convergence condition can only
be satisfied at future infinity where the scale factor is proportional
to $\tau$. The fact that quantum effects are essentially responsible
in allowing the existence of such a geometry makes the violation
of the convergence condition only more plausible. Recall that 
in four-dimensional (4D) scenarios the violation of the energy conditions
is often attributed to `quantum' matter \cite{BD}.


We now discuss the distribution of quantum matter and the status of the energy
conditions in our solutions. As mentioned above, in our
model the total matter stress tensor is determined solely by the one-loop 
contribution $\langle T_{\mu\nu} \rangle$, which in turn depends on 
the choice of the matter state as defined in (\ref{qt}). 
Since it is the first branch that gives an interesting cosmological evolution,
we calculate $\langle T_{\mu\nu} \rangle$  only for this case. The only 
nonvanishing component 
turns out to be
$\langle T_{\tau \tau} \rangle = \kappa / 4\left[ 3\ddot{a} /a
- 8 \left( \dot{a} /a \right)^{2} \right]$.
For the first branch solution (\ref{a1st}), this gives
\begin{equation}
\label{VEVtau}
\langle T_{\tau \tau} \rangle = -{\kappa \alpha^2 \over 4 [ (\alpha \tau
+ \beta )^2 +\kappa ]^{3/2}} \left[ 3 (\alpha \tau + \beta ) + 5\sqrt{ 
( \alpha \tau + \beta )^2 + \kappa} \right]
\ \ ,
\end{equation}
which is always nonpositive. To physically interpret the matter state, note
that the above expression vanishes asymptotically as $\tau \to \pm \infty$.
Since for this branch the scalar curvature also vanishes in these regimes, it
implies that matter state is the `in' or `out' Minkowski vacuum. 

The Weak Energy Condition (WEC) is violated in this branch since the 
average density of matter is negative. However, unlike in 4D
Einstein gravity, in this 2D model the equations of motion show that the 
Ricci tensor is not exclusively determined by the matter stress tensor.
Thus, just the fact that $\langle T_{\tau \tau} \rangle$ is nonnegative
does not necessarily guarantee defocusing of the timelike geodesics and, 
hence, the cure to the graceful exit problem. Nevertheless, as shown above, 
explicit calculation of the Ricci scalar for the solution (\ref{a1st}) does 
bear out such a behavior of geodesics in these spacetimes. The effect of 
violation of energy conditions on the solution to the graceful exit problem
in 4D is studied in Ref. \cite{BM}.

We conclude the paper with a few remarks. First, it is important to note that
in Rey's solution to the graceful exit problem the barrier $N < 24$ is not
merely a restriction on the permissible number of matter fields in that 
theory. More important, it violates the large $N$ approximation \cite{CGHS}, 
which is assumed at the outset when only the one-loop correction is retained 
in the effective action after expanding it in powers of 
the Planck's constant. In fact, large $N$ approximation is used to drop
the higher order corrections as well as corrections due to ghost 
contributions. Therefore, in order to get a consistent model one should 
allow for $N$ to be large in its solutions.

Second, here we have concentrated only on 2D cosmological models. 
However, realistically it is important to address the 
graceful exit problem in 4D. The no-go theorems have shown that modifying the 
4D tree level action by adding realistic dilatonic potentials 
\cite{Brus-nogo}, axions \cite{KMO} or even tachyons \cite{MMS} does not 
cure this problem. Calculations within a quantum cosmology
setting do indicate the possibility of an exit \cite{QC}. Also the result that
an exit is possible only if the weak energy condition is violated indicate
that the graceful exit problem might be a classical feature that can be
cured only by including quantum corrections. This urges one to study 
solutions of the 4D effective action (see, eg., \cite{Ven-rev})
expanded to the next higher order in the
coupling, which is a nontrivial problem. An alternative is to 
perform a semiclassical treatment by quantizing only the matter fields and
retaining only one-loop terms by invoking the large $N$ approximation. This
option is not trivial either because of the presence of higher derivative 
terms. Thus, for a beginning, a tractable calculation that includes quantum
effects can be outlined as follows. In the spirit of the minisuperspace
program, one imposes isotropy at the level of the vacuum 4D 
tree action and performs an S-wave reduction (by integrating over the angular
coordinates). The resulting 2D action is the CGHS action. Thus the 
semiclassical solutions of the CGHS action, which we have discussed in this
paper, illustrate the semiclassical behavior of the S-wave sector of the
4D theory. This suggests that perhaps the 4D effective action expanded to 
higher orders may be devoid of the graceful exit problem.

Finally, note that the Polyakov-Liouville term in the one-loop corrected 
action breaks both conformal invariance and SFD. Since the 4D effective action
preserves SFD at higher orders in the perturbative expansion \cite{Meiss}, 
one might question if our results, which break SFD, have any bearing on the 
4D quantum-corrected solutions. We do not
have a completely satisfactory answer to this question, except to point out
that the large $N$ approximation allows us to neglect the quantum corrections
due to the dilaton, metric, and ghosts. Making such an approximation on the
S-wave sector of the 4D effective action expanded to first order will 
presumably result in a theory without SFD. It is in this limit that our
2D solutions might make a correspondence with their 4D counterparts.

The authors gratefully acknowledge financial support from the Inter--University
Centre for Astronomy and Astrophysics, Pune, India.

\vfil 
\pagebreak

\centerline{\bf FIGURE CAPTIONS}
\vspace{.2in}
{\bf Fig. 1} The scale factor $a(\tau)$ versus $\tau$ as given in 
Eqn. (\ref{a1st}) (+ sign). Here $\alpha =1$, $\beta =0$ and $\kappa =2$.
 
{\bf Fig. 2 (a)} $\tau$ as a function of $\xi = e^{-2\phi}$. Here,
$\kappa =2$.
 
{\bf Fig. 2 (b)} The scale factor $a(\tau)$ versus $\tau$ as given
(in parametric form) by Eqns. (\ref{a2}) and (\ref{tau2}). Here, $\kappa =2$.

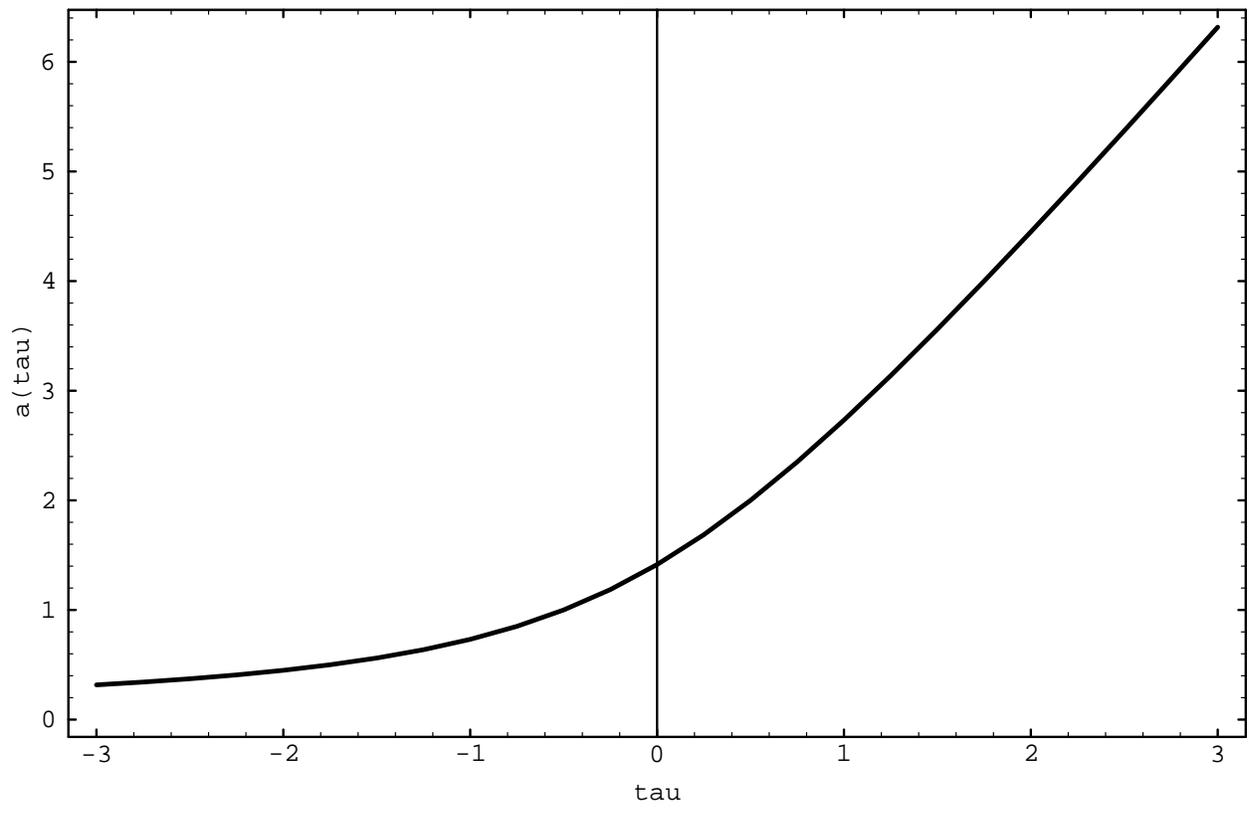
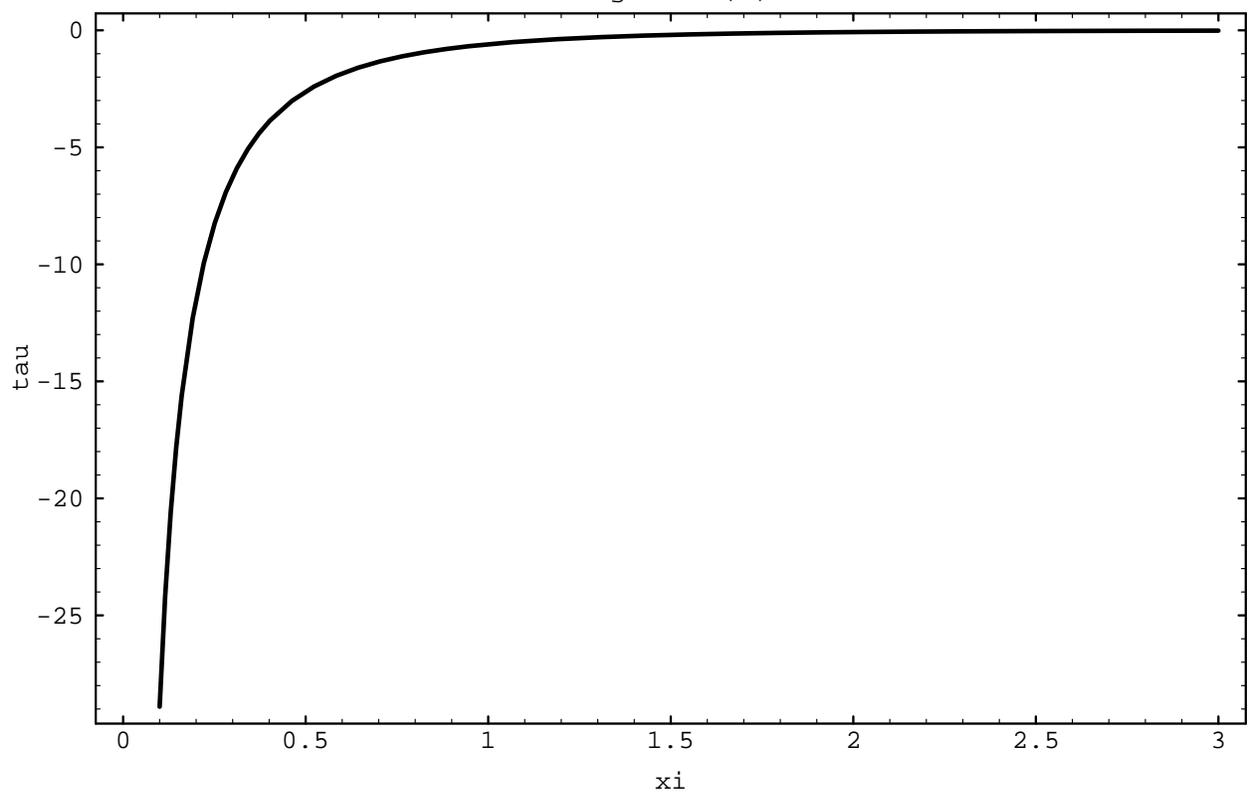
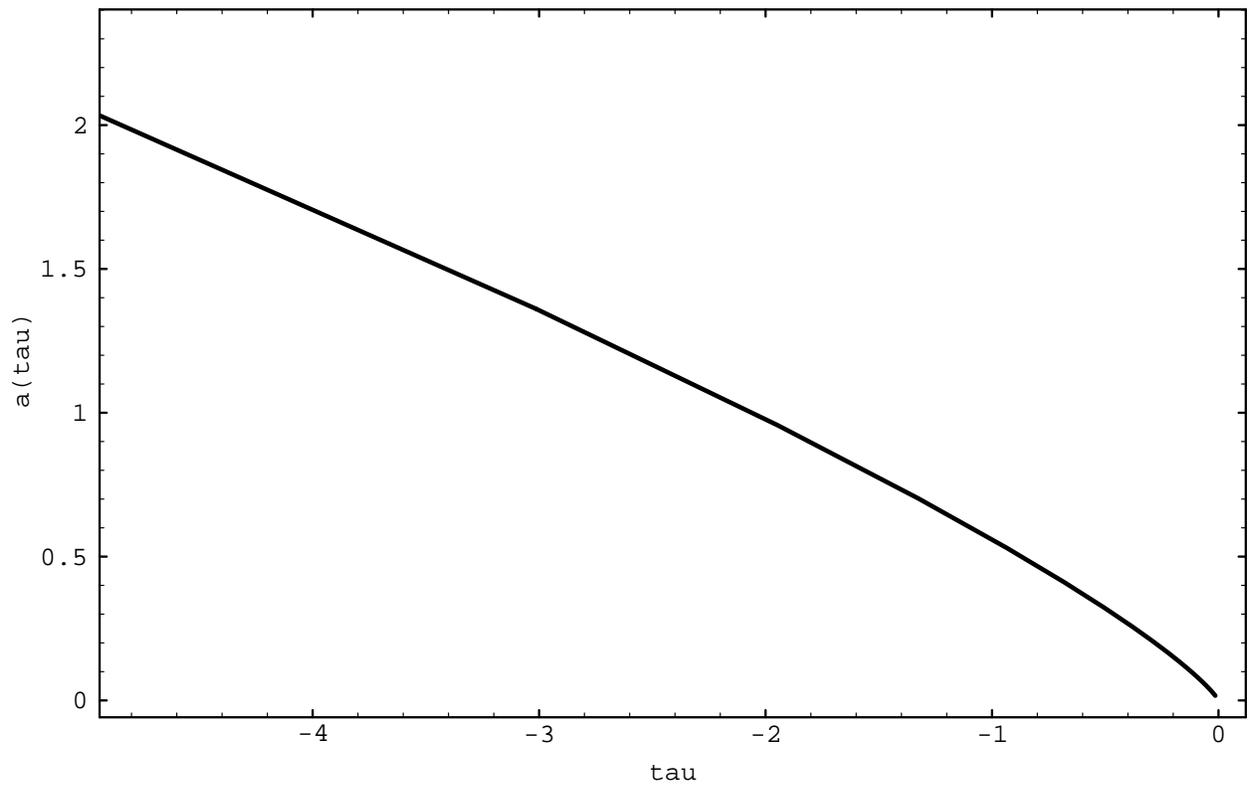
\end{document}